# The Source of Solar Energy, ca. 1840-1910:
# From Meteoric Hypothesis to Radioactive Speculations

Helge Kragh[*]

**Abstract**: Why does the Sun shine? Today we know the answer to the question and we also know that earlier answers were quite wrong. The problem of the source of solar energy became an important part of physics and astronomy only with the emergence of the law of energy conservation in the 1840s. The first theory of solar heat based on the new law, due to J. R. Mayer, assumed the heat to be the result of meteors or asteroids falling into the Sun. A different and more successful version of gravitation-to-heat energy conversion was proposed by H. Helmholtz in 1854 and further developed by W. Thomson. For more than forty years the once so celebrated Helmholtz-Thomson contraction theory was accepted as the standard theory of solar heat despite its prediction of an age of the Sun of only 20 million years. In between the gradual demise of this theory and the radically different one based on nuclear processes there was a period in which radioactivity was considered a possible alternative to gravitational contraction. The essay discusses various pre-nuclear ideas of solar energy production, including the broader relevance of the question as it was conceived in the Victorian era.

## 1  Introduction

When Hans Bethe belatedly was awarded the Nobel Prize in 1967, the presentation speech was given by Oskar Klein, the eminent Swedish theoretical physicist. Klein pointed out that Bethe's celebrated theory of 1939 of stellar energy production had finally solved an age-old riddle, namely "how it has been possible for the sun to emit light and heat without exhausting its source."[1] Bethe's theory marked indeed a watershed in theoretical astrophysics in general and in solar physics in particular.

---

[1] https://www.nobelprize.org/nobel_prizes/physics/laureates/1967/press.html (accessed July 2016).

Thanks to Bethe and his followers we know today in great detail that the Sun is a huge fusion reactor. Although the solution of the riddle may be safely dated to 1939, more than a decade earlier Arthur Eddington had anticipated that thermonuclear reactions power the Sun [Hufbauer 2006]. Our present knowledge of solar energy production started with either Eddington or Bethe, but it should not make us forget that there were many earlier attempts to solve the riddle. This paper surveys the attempts during the period from about 1840 to 1910, shortly before the atomic nucleus became a reality. From a historical point of view this earlier and unsuccessful period is no less interesting than the later success story.

      The concept of the Sun changed dramatically in the mid-nineteenth century as a consequence of the discovery of the fundamental laws of thermodynamics. The Sun was now conceived as a huge heat engine and the question of the source of its heat and light became of central importance to the emerging fields of solar physics and astrophysics. Like any other heat engine the Sun was in need of fuel, and when the fuel ran down it would supposedly cease to shine. On the top of that, the second law of thermodynamics implied that the emitted energy would increasingly become less useful and eventually the solar system, and indeed the whole stellar universe, would end in a lifeless equilibrium state with no possibility of recovery. In any case, the message of the new energy physics was depressingly clear: The Sun was irreversibly on its way towards extinction.

      The first mechanical theories of solar heat were meteoric, based on the assumption of massive meteor showers falling into the Sun and thus feeding it with energy. However, it was soon realized that meteors were unable to generate the necessary energy and since about 1860 another theory became widely accepted. According to the Helmholtz or Helmholtz-Thomson theory, gravitational contraction was the source of the Sun's thermal energy. The theory predicted a solar lifetime somewhere in the interval between 10 and 100 million years. Although the contraction theory was supported by most physicists and astronomers, there was no shortage of alternatives promising an almost eternal life to the solar system. Throughout the period the discussion concerning the Sun's life and death was closely connected with the question of the age of the Earth and its habitability in the future. With regard to this question, the physicists' calculations disagreed violently with the geologists' estimate of the Earth's age [Burchfield 1975]. During the first decade of the twentieth century a new energy source aroused much interest, namely radioactivity and more generally subatomic energy. But the interest was short-lived

and by the end of the 1910s the problem of the Sun's energy remained unsolved. And yet there was light at the end of the tunnel.

## 2  The dying Sun

In his classic *Treatise on Astronomy*, a work dating from 1833, the distinguished British astronomer and natural philosopher John Herschel briefly dealt with the question of the source of the Sun's energy. We are "completely at a loss," he admitted [Herschel 1833, p. 212]. A few decades later the scientists were still at a loss, but not quite as much. An answer to the question seemed within reach. The conception of the Sun changed materially in the 1840s as a consequence of the discovery of the principle of conservation of energy. While previously it had made sense to conceive the Sun as an eternal cosmic object in agreement with the uniformitarian methodology of Charles Lyell and others, it now became natural, and indeed imperative, to discuss the source of the Sun's heat as well as the time at which it would cease to shine. As the American astronomer Simon Newcomb [1878, p. 505] pointed out, it was only with the discovery of energy conservation that these difficult questions were recognized to belong to the realm of science.

     Latest by the 1860s the dying Sun had become a reality to the Victorian mind whether scientific or literary [Gold 2010]. The dire consequences of the new laws of thermodynamics were expressed by Lord Tennyson, the celebrated poet, who in 1849 wrote:

> The stars, she whispers, blindly run;
>   A web is wov'n across the sky;
>   From out waste places comes a cry,
> And murmurs from the dying Sun.

And in a later poem of 1886:

> Dead the new astronomy calls her, …
> Dead, but how her living glory lights the fall, the
>   dune, the grass!
> Yet the moonlight is the sunlight, and the sun
>   himself will pass.

## 2.1 Laws of thermodynamics

The fundamental insight that all natural processes are governed by an indestructible force called energy, and that heat is but one form of the more general concept of energy, was first formulated by the German physician Julius Robert Mayer in 1842 [Lindsay 1973; Caneva 1993]. His paper appeared in *Annalen der Chemie und Pharmacie*, a journal edited by the great chemist Julius von Liebig. Slightly later the Manchester brewer and natural philosopher James Prescott Joule independently announced his version of energy conservation to the 1843 meeting of the British Association for the Advancement of Science. From some rather inaccurate experiments Mayer calculated the mechanical equivalent of heat to 1 cal = 3.65 J, while Joule obtained the substantially better value 1 cal = 4.24 J (the modern value is 4.184 J). Several other scientists were involved in the complex discovery history, a classic case of "simultaneous discovery" [Kuhn 1969; Elkana 1974]. Among the secondary co-discoverers were the Danish physicist L. August Colding, the French engineer Marc Séguin, and the British lawyer and physicist William Grove.

     Neither the work of Mayer nor that of Joule aroused much immediate attention, and it took a couple of years until the law of energy conservation was fully digested and formulated in a precise and comprehensive manner. Such an understanding was presented in *Über die Erhaltung der Kraft* (On the Conservation of Force), a most important memoir of 1847 written by the 26-year-old German physician and physicist Hermann von Helmholtz. Remarkably, his masterpiece was not accepted for publication in *Annalen der Physik und Chemie* and consequently it appeared as a privately printed booklet. Helmholtz's memoir not only provided a mathematical formulation of the law of energy conservation, it also demonstrated the unifying role of energy or "force" in all areas of natural philosophy. Helmholtz stressed that energy conservation was an empirical generalization, neither a philosophical doctrine nor a mere verbal truism as some of his contemporaries thought. John Herschel was one of those who thoroughly misunderstood the meaning of the law [James 1985].

     It took only a decade until the law of energy conservation, alias the first law of thermodynamics, was supplemented with another and no less fundamental law. In a paper of 1850 the German physicist Rudolf Clausius, at the time professor at the Royal Artillery and Engineering School in Berlin, formulated a synthetic theory that rested on two fundamental principles of what soon became known as

"thermodynamics." The word was coined by the Irish-born Scotsman William Thomson four years later. The first of Clausius' principles was energy conservation and the second was the statement that it is impossible for a self-acting cyclic machine to convey heat from a body of lower temperature to another at a higher temperature. Thomson – who is better known as Lord Kelvin or just Kelvin (he was ennobled in 1892) – said about the same in a paper of 1851. The following year, after having read Helmholtz's memoir, he formulated the second principle as an inbuilt and universal tendency in nature towards the dissipation of energy. He added: "Any restoration of mechanical energy, without more than an equivalent of dissipation, is impossible in inanimate material processes" [Thomson 1882, p. 514].

Clausius continued to reformulate and generalize the second law, and in 1865 he introduced the concept of "entropy" as a measure for the growing dissipating force. According to Clausius, the entropy difference between two physical states A and B was given by

$$\Delta S = \int_A^B \frac{dQ}{T},$$

where $T$ is the absolute temperature and the path of integration corresponds to a reversible transformation from A to B; the quantity $dQ$ denotes an infinitesimal amount of heat absorbed by the system. As he emphasized, the entropy of an isolated physical system will inevitable tend towards a maximum. Thomson always preferred to speak of dissipation of heat or energy, a concept he found more easy to use and to visualize than the abstract notion of entropy. To him, dissipation referred to a situation in which two bodies at different temperatures placed in contact transfer heat from the warmer to the colder without work being done. Although the energy is conserved, it has become dissipated in the sense that the system's capacity to perform mechanical or other work has diminished.

Whether in Clausius' or Thomson's version, by the late 1850s thermodynamics was practically complete and its significance widely recognized. In his presidential address of 1863 to the British Association, the chemist William Armstrong [1864, p. lx] called the dynamical theory of heat "probably the most important discovery of the present century." The high appreciation was shared not only by many physicists and chemists but also by a growing number of astronomers. Indeed, the solar system played an important role to several of the pioneers of thermodynamics who thought that the new science required reconsideration of the Sun and its sources of energy.

## 2.2 Mayer's meteorite hypothesis

As early as 1841, slightly before his first publication on energy conservation, Mayer contemplated the idea that the Sun's heat might be due to the capture of meteors from the surrounding space. Five years later he had a paper ready on the subject, but it was only published in 1848 and then on his own expense.[2] Mayer considered various sources for the solar heat production, including the possibility that it was due to the burning of coal or some other chemical combustion process. However, in 1838 the French physicist Claude Servais Pouillet, director of the Conservatoire des Arts et Métier in Paris, had determined what he called the "solar constant" by means of a pyrheliometer, an invention of his own. Describing the new physical constant as "the essential element in the Sun's constant calorific power" he reported its value to be 1.7633 cal cm$^{-2}$ min$^{-1}$ [Pouillet 1838; Kidwell 1981]. Aware of Pouillet's work Mayer realized that chemical processes were hopelessly inadequate to fuel the Sun. He also considered the Sun's rotation as an energy source, but only to reject the possibility.

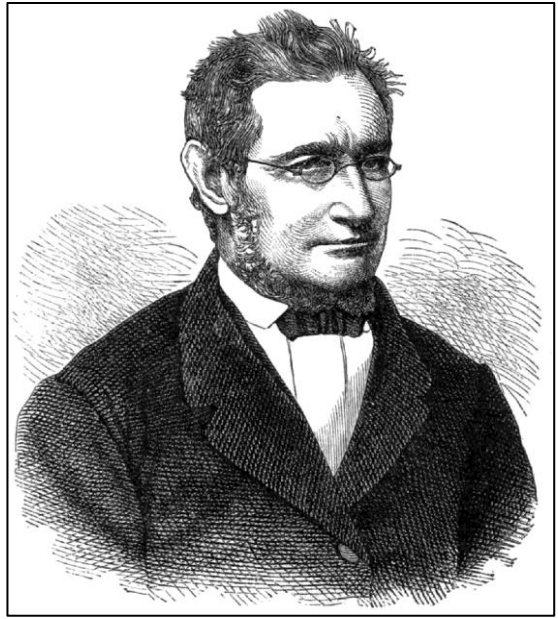

**Fig. 1**. Julius Robert Mayer (1814-1878). Source: https://upload.wikimedia.org/wikipedia/commons/2/2b/Julius_Robert_von_Mayer.jpg

---

[2] Mayer's privately printed paper entitled "Beiträge zur Dynamik des Himmels" (Contributions to Celestial Dynamics) was translated into English in 1863 and again in 1865. I quote from the translation in Lindsay [1973, pp. 147-195].

Solar activity, Mayer concluded, must ultimately be gravitational in origin, namely, the result of a constant falling of meteors or asteroids into the Sun. This he considered "the only tenable explanation of the heat of the sun." Estimating the velocity of an asteroid when striking the Sun to be between 446 and 630 km/s, Mayer wrote:

> An asteroid, therefore, by its fall into the sun developes [sic] from 4600 to 9200 times as much heat as would be generated by the combustion of an equal mass of coal. … We find the heat developed by the asteroid to be from 7000 to 15,000 times greater than that of the oxyhydrogen mixture. From data like these [we infer] … that no chemical process could maintain the present high radiation of the sun.

The term "oxyhydrogen mixture" was a reference to the exothermic synthesis of gaseous water as given by $2 H_2 + O_2 \rightarrow 2 H_2O + 572$ kcal/mole.

Rather than assuming meteors to fall in straight lines from space towards to Sun, Mayer argued that their orbits were spirals, a result due to the resisting ethereal medium. "Scientific men do not doubt the existence of such an aether," he ascertained, evidently counting himself among the scientific men. Contrary to Helmholtz, Thomson and their followers, Mayer did not conclude that the Sun's activity would eventually cease. His aim was to describe a realistic steady state scenario in which the radiated energy was compensated by a constant energy supply provided by falling meteors. Not only was the energy of his Sun in a steady state, so was its mass and that in spite of its continual absorption of meteoric matter. Mayer calculated that each minute there would fall upon the Sun a mass between $94 \times 10^{12}$ and $188 \times 10^{12}$ kg. He realized that the increased mass of the Sun would result in a shortening of the sidereal year of the order half a second, an effect disagreeing with observations.

Not willing to abandon his steady state solar model, he came up with a speculative hypothesis relying on the ether. At the time the wave theory of light was firmly established, and Mayer knew that according to this theory light was weightless, so how could the emission of sunlight compensate for the meteoric mass increase?[3] His answer was to conceive the ether as a corpuscular substance and in this way to arrive at the conclusion that "the sun, like the ocean, is constantly losing and receiving equal quantities of matter." The increase in the Sun's volume was less problematic, for Mayer estimated that it would be imperceptible to astronomical observation. It would take about 50,000 years until the apparent diameter had increased by one arc second. Mayer's 1848 essay on solar dynamics was his last scientific paper. Personal tragedies in that year as well as the failed appreciation of his work affected him deeply, and after a failed suicide attempt in 1850 he was committed to a mental institution. Although he was eventually released and able to pursue his scientific interests, he never fully recovered.

A meteoric theory of solar heat was independently suggested by the Scottish engineer and physicist John James Waterston, who read a paper on the subject at the 1853 meeting of the British Association. Unaware of Mayer's earlier work (which was only translated into English in 1863), Waterston suggested that the Sun's heat originated in the influx of a large number of meteors; he thought that these came mainly from outside the solar system and hit the Sun perpendicularly to its surface. He estimated that, if the Sun's heat was entirely due to meteors, its radius would increase by approximately five metre per year. Whereas Mayer was concerned with the problem of the Sun's increased mass and its astronomical consequences, Waterston either disregarded it or was just unaware of it. However, he mentioned it in a paper of 1860 in which he gave various estimates of the Sun's excessive temperature as caused by the capture of very large meteors [Waterston 1860].

Waterston's early speculations did not result in a scientific paper, but they were known to and much appreciated by Thomson, who soon developed what he called Waterston's "remarkable speculation of cosmical dynamics" into a quantitative meteoric theory of the Sun's heat [James 1982; Burchfield 1975, pp. 22-31]. Waterston prepared a manuscript on his prescient ideas of the kinetic theory of heat and solar energy as early as 1845 and submitted it to the Royal Society of London, which however rejected it for publication [Waterston 1892]. On the initiative

---

[3] Mayer's speculations were anticipated in the eighteenth century, where the generally accepted corpuscular theory of light, assuming the light particles to carry mass, seemed to imply a continual *decrease* of the Sun's mass. Benjamin Franklin and Joseph Priestley were among those who responded to the problem [Cantor 1983]. One answer was to deny any loss of matter at all, either by denying the corpuscular theory in favour of a wave theory of light or by postulating that comets and other celestial bodies fell into the Sun and thus refuelled it.

of Lord Rayleigh it was eventually published in 1892, many years after Waterston had passed away.

## 2.3  Thomson and the Sun's heat

In a paper of 1854 with the title "On the Mechanical Energies of the Solar System," Thomson [1882, pp. 1-25] presented his own version of the meteoric-gravitational hypothesis. Like Waterston, he was still unaware of Mayer's theory. There were only three conceivable theories for the Sun's heat, Thomson said, and two of them he dismissed as quite untenable. One was the idea that the Sun was an originally heated body which gradually was losing its heat; and the other was that its heat was due to chemical actions. Transformation of gravitational energy into heat by means of falling meteors was the only acceptable explanation of the solar heat problem. "According to this form of the gravitation theory," he wrote, "a meteor would approach the Sun by a very gradual spiral … until it begins to be more resisted, and to be consequently rapidly deflected towards the Sun; then the phenomenon of ignition commences; after a few seconds of time all the dynamical energy the body had at the commencement of the sudden change is converted into heat and radiated off; and the mass itself settles incorporated in the Sun."

As to the necessary influx of cosmic matter into the Sun, Thomson found that it amounted to 100 Earth masses in the course of 4750 years. Large though this amount of matter was, he claimed that it was "not more than it is perfectly possible does fall into the Sun." The meteors considered by Thomson differed from Mayer's asteroids in the sense that they were circulating round the Sun far inside the orbit of the Earth; he explicitly denied the hypothesis based on impacts of celestial bodies beyond the solar system.

Thomson believed that the Sun's rotation upon its axis was caused by the same mechanism that was responsible for its emission of heat, the falling meteoric matter. From this hypothesis he reasoned that the Sun could scarcely have shone for much more than 32,000 years and that it was unlikely to maintain its present activity for much more than 300,000 years in the future. This was the first attempt to calculate on the basis of physical theory the age of the Sun and the end of its life. To the mind of Thomson, the question of the Sun's heat related as much to the second law of thermodynamics as to the first law. Concerning the radiation energy emitted by the Sun he wrote in his 1854 paper that it "is dissipated always more and more widely through endless space, and never has been, probably never can be, restored to the Sun." Thomsen further connected his theory to paleoclimatology, stating that "The meteoric theory affords the simplest explanation of past changes of climate on the earth."

Although confident that his meteoric theory was essentially correct, Thomson realized that it lacked verification. When the French astronomer Urbain Leverrier in 1859 announced the famous Mercury perihelion anomaly – namely that the advance of the planet's perihelion disagreed with Newton's law of gravity – for a brief while Thomson thought that the discovery provided evidence for his theory. Leverrier's favoured explanation of the discrepancy was to hypothesize the existence of intra-Mercurial lumps of matter, and might this matter not be meteors slowly on their way towards the Sun? However, Thomson's calculations soon showed that the mass of meteors necessary to supply the Sun's heat did not agree with the intra-Mercurial mass that might explain the observed anomaly. Moreover, if a dense meteoric cloud existed within Mercury's orbit it would cause problems for the passage of comets. What initially had seemed to be evidence for the meteoric hypothesis was not, after all, evidence. The Mercury anomaly persisted until Einstein famously solved it in 1915 on the basis of his general theory of relativity. The Mayer-Waterston-Thomson theory was short-lived as it presupposed an extraordinary number of meteoric bodies currently falling into the Sun, an assumption for which there was no evidence at all. By the 1870s the meteoric theory was practically dead as it was recognized that the part of the solar heat that could be ascribed to the influx of meteors was negligibly small.

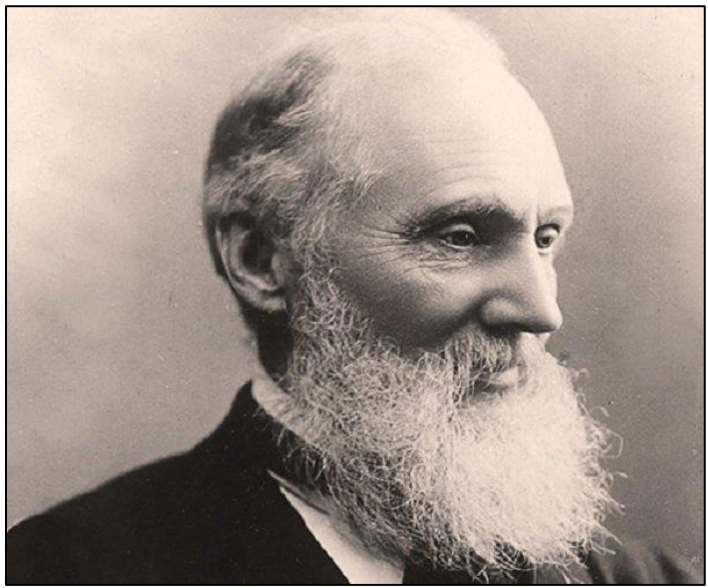

**Fig. 2.** William Thomson, alias Lord Kelvin (1824-1907). Image credit: The famous people website. http://www.thefamouspeople.com/profiles/images/lord-kelvin-9.jpg

By the fall of 1854 Thomson had become aware of Helmholtz's "most interesting popular lecture" in which the German scientist proposed an alternative gravitational theory for the Sun's heat (see Section 3.2). But Thomson [1884, pp. 34-39] did not endorse Helmholtz's contraction mechanism which he thought was insufficient to deliver the necessary amount of heat. He maintained that the Sun must be constantly powered by "the mechanical action of masses coming from a state of very rapid motion round the sun to rest on his surface." In 1862 Thomson finally acknowledged Mayer's priority, yet at the same time arguing that the Sun was presently a cooling body with no appreciable energy compensation in the form of heat generated by the influx of meteors. He now abandoned his original meteoric theory and instead adopted a version inspired by Helmholtz. In an article entitled "On the Age of the Sun's Heat" published in *Macmillan's Magazine*, Thomson [1891, pp. 356-375] accepted gravitational contraction as the true source of the Sun's heat. However, he still thought in terms of a meteoric theory, but now conceived the energetic actions of the meteors to belong to the distant past. "That *some form* of the meteoric theory is certainly the true and complete explanation of solar heat can scarcely be doubted," he wrote (emphasis added). On the basis of rough calculations Thomson not only suggested limits for the age of the Sun, but also for its future emission of energy. He concluded that it was "on the whole most probable that the sun has not illuminated the earth for 100,000,000 years, and almost certain that he has not done so for 500,000,000 years." And this was not all:

> As for the future, we may say, with equal certainty, that inhabitants of the earth can not continue to enjoy the light and heat essential to their life, for many million years longer, unless sources now unknown to us are prepared in the great storehouse of creation.

That there were indeed such unknown sources of energy only became evident some forty years later.

In the same paper Thomson briefly confronted Charles Darwin's new and controversial theory of biological evolution with the physically based time-scale of the Earth, rhetorically asking if geological evidence for the age of the Earth was really more trustworthy than the evidence based on calculation and sound physical reasoning. This was the beginning of a famous controversy concerning the age of the Earth and not primarily the age of the Sun [Burchfield 1975]. Thomsen's paper of 1862 was reprinted as an appendix to his and P. G. Tait's widely read textbook *Treatise of Natural Philosophy* (1867) and in this way became known to almost all British physicists.

## 3   The Helmholtz-Thomson contraction theory

What in the late Victorian era was celebrated as the authoritative and probably correct theory of solar energy production was a gravity-to-heat theory, but not in the meteoric version with roots in Mayer, Waterston and Thomson. When Agnes Clerke [1893, p. 378], an astronomer and chronicler of astronomy, wrote on "the theory of solar energy now generally regarded as the true one," she was referring to Helmholtz rather than Thomson. And yet the theory was often named the Helmholtz-Thomson (or Helmholtz-Kelvin) theory. The essence of this theory was that the Sun itself contracted slightly and the gravitational energy released during its formation manifested itself in the liberation of light and heat. Although Mayer was well acquainted with the idea of gravitational contraction and applied it to the internal heat of the Earth, he did not consider it a possible mechanism for explaining the heat of the Sun.

### 3.1   Kantian preludes

The contraction theory can be traced back to a little known essay in the *Berlinische Monatsschrift* that the famous philosopher Immanuel Kant wrote in 1785 and in which he critically and competently reviewed various ideas concerning the nature and origin of lunar craters. According to Kant, most of these ideas were quite wrong as they relied on the assumption of volcanic activity on the Moon. As he saw it, the craters were scars on the Moon's surface dating from the time it was formed from the gravitational condensation of a primitive gaseous substance. In the course of arguing for this hypothesis, Kant [2015, p. 425] was brought to consider the issue of the origin of solar heat, which he thought could be understood in purely scientific terms. As he wrote:

> In the case of a natural phenomenon such as the heat of the Sun, … I think it unacceptable to come to a halt and in desperation invoke an immediate divine decree as an explanation. This latter must admittedly form the conclusion of our investigation when we talk of nature as a whole; but in every epoch of nature, since no one of them can be shown by direct observation to be absolutely the first, we are not relieved of the obligation to search among the causes of things as far as is possible for us, and follow the causal chain in accordance with known laws as far as it extends.

Kant based his qualitative explanation of the Sun's heat on the cosmic nebular hypothesis he had suggested thirty years earlier and according to which the primeval universe was a chaotic distribution of particles governed by attractive and repulsive forces. Gravity would slowly lead to solar systems formed from the original chaos [Kant 1981]. Ideas somewhat similar to Kant's were developed later in the century by William Herschel in England and Pierre-Simon Laplace in France, and they eventually became known as the nebular hypothesis or the Kant-Laplace hypothesis [Brush 1987].

The novelty of Kant's essay of 1785 was that it suggested a mechanism for the production of the high temperatures of the stars formed from a cold cloud of gaseous material. How did the "element of heat" become concentrated in the stars? Kant wrote: "If it be assumed … that the original matter of all celestial bodies, in the whole vast space in which they now move, was initially distributed in gaseous form, and was formed initially in accordance with the laws of chemical attraction, then Crawford's discoveries suggest how the formation of celestial bodies [is linked with] the production of the requisite enormous degrees of heat." The Irish-born physician and chemist Adair Crawford was a pioneer in measurements of the specific heats of gases and he had recently observed that when a gas was compressed, heat would evolve. Picking up on Crawford's discovery Kant argued that gravitational compression was the source of the intense heat of the Sun and the other stars. The planets too would be heated in this way, but Kant thought that the amount of the created heat element would be proportional with the mass of the body and hence be important only for the Sun.

Kant's innovative attempt to explain the solar heat, based on the old conception of heat as a caloric fluid, seems to have been forgotten. Its relevance for the later discussion of solar heat was first pointed out by the American geologist Georges Ferdinand Becker [1898]. It took sixty years until similar ideas were suggested on the basis of the new law of energy conservation and conversion. In his ill-fated manuscript of 1845 Waterston not only proposed a meteoric hypothesis of the Sun's heat, in a note he also suggested that the gravitational contraction of the Sun might be a source of its heat. Referring to the "Nebular Hypothesis of Laplace" he wrote that "The intense activity of the molecules of the Sun's mass may be viewed as the result of, or to have been originally produced by, its centripetal force while condensing." Moreover, he came up with a figure of the relationship between the Sun's contraction and the solar radiation produced by it [Waterston 1892, p. 55; Tassoul and Tassoul 2004, p. 71]:

> During one year the solar force upon a square foot at the Earth's mean distance from the Sun is equal to … one ton raised one mile per day… If the Sun is supposed to contract uniformly throughout its mass so that it radius becomes 3 ⅓ miles less in consequence of the general increase of density, the force generated is sufficient to supply the solar radiation for about 9000 years.

Although Waterston's 1845 paper was not published, he referred to similar ideas in his communication to the 1853 meeting of the British Association which Helmholtz attended. It is unclear if Waterston thought that the Sun's heat was primarily due to contraction or to meteoric falls, and he may have considered the two processes to occur simultaneously rather than being alternatives.

## 3.2   Helmholtz, solar energy and the nebular hypothesis

On 7 February 1854 Helmholtz gave a broad-ranging lecture on "The Interaction of Natural Forces" in Königsberg, the city where Kant had lived his entire life and acted as a professor of philosophy. The main subject was the law of energy conservation, but he also discussed the second law and the nebular hypothesis. Helmholtz emphasized the cosmic consequences of what he supposed to be the universally valid second law of thermodynamics, namely the "heat death" (*Wärmetod*) that followed from it. The heat death scenario for the solar system had earlier been suggested by Thomson, but Helmholtz [1995, p. 30] was the first to extend it to a cosmological scale: "If the universe be delivered over to the undisturbed action of its physical processes, all force will finally pass into the form of heat, and all heat come into a state of equilibrium." What he called the "day of judgment" would occur when the Sun lost its light and high temperature. His lecture in Königsberg became more widely circulated after it was translated into English in *Philosophical Magazine* in 1856.

"It may be calculated," Helmholtz said, that "if the diameter of the sun were diminished only the ten-thousandth part of its present length, by this act a sufficient quantity of heat would be generated to cover the total emission for 2,100 years." Although he may have known of Waterston's estimate from the 1853 meeting of the British Association, he did not refer to him in his lecture. He also did not refer to Kant's paper of 1785, probably because it was unknown to him. On the other hand, he praised Kant's cosmological treatise of 1755 and "the depth to which he [Kant] had penetrated into the fundamental ideas of Newton." Meteors played a crucial role in the theories of Mayer and Thomson, but their role was much more peripheral in Helmholtz' theory. Here the role of meteors was limited to evidence for the ongoing evolutionary processes in the solar system.

For the potential gravitational energy of the Sun, supposed to be of uniform density $\rho$ and with mass $M$ and radius $R$, Helmholtz gave the expression

$$V = -\frac{3}{5}\frac{r^2 M^2}{Rm}g$$

Here $m$ denotes the mass of the Earth and $r$ its radius, and $g$ is the gravitational acceleration at the Earth's surface [Stinner 2002; Tort and Nogarol 2011]. Using the relation between $g$ and Newton's universal constant of gravitation $G$,

$$g = \frac{Gm}{r^2},$$

the expression can be stated in the more familiar and compact form

$$V = -\frac{3}{5}\frac{GM^2}{R}$$

In the later textbook version the result is derived by conceiving the spherical Sun as made up of concentric thin shells of width $dr$ and mass $d\mu = 4\pi r^2 \rho dr$. The mass contained within the radius $r$ is

$$\mu(r) = \frac{4}{3}\pi r^3 \rho,$$

and the potential gravitational energy thus

$$V = -G \int_0^R 4\pi r^2 \rho \frac{\mu(r)}{r} dr = -\frac{16}{15} G\pi^2 R^5$$

Inserting $4\pi R^3 \rho = 3M$ yields the above result. Helmholtz stated the heat energy corresponding to a temperature increase $\Delta T$ as

$$E = AgMc\Delta T,$$

where $c$ is the specific heat capacity and $Ag$ "represents the mechanical equivalent of the unit of heat." By equating $E$ and the numerical value of $V$, he found

$$\Delta T = \frac{3}{5} \frac{r^2 M}{ARmc}$$

On the arbitrary assumption that the specific heat capacity of the Sun is the same as that of water (1 cal g⁻¹ per degree C), Helmholtz arrived at

$$\Delta T = 28{,}611{,}000 \; °C$$

or roughly 28 million degrees. He realized that the Sun was probably denser at the centre than near the surface and that this correction would result in an even greater amount of heat. If the density varies as

$$\rho = \rho_0 r^{-2},$$

the heat production will increase by a factor of 5/3 [Poincaré 1911, p. 202].

Although Helmholtz did not give a specific estimate of the Sun's age based on the gravitational mechanism, he stated that if $R$ was diminished by a ten-thousandth part of its present value it would generate an increase in the Sun's temperature of 2,861 °C. "And as, according to Pouillet, a quantity of heat corresponding to 1¼ degree is lost annually in such a mass, the condensation referred to would cover the loss for 2,289 years." That is, a total contraction at constant radiation rate would take approximately 23 million years.

According to the virial theorem the average kinetic energy equals one half the average potential energy. It follows that only half the change in gravitational potential energy is available to be radiated away as the Sun shrinks. The radiation energy will thus be

$$V_{rad} = \frac{3}{10} \frac{GM^2}{R},$$

and the same amount will appear as thermal energy in the Sun. Helmholtz did not refer to this correction, probably because he was unaware of the virial theorem which, in its modern version, was only introduced by Clausius in a lecture of 1870.

Helmholtz was instrumental in calling attention to Kant as a great natural philosopher and in linking together the philosopher's view with that of Laplace, the mathematician and astronomer, into the Kant-Laplace nebular hypothesis or world view.[4] To the mind of Helmholtz, the question of the Sun's heat and age belonged to the framework of the nebular hypothesis and he regarded the latter as a precondition for answering the first. This came to be the accepted view, such as illustrated by the evaluation offered by Clerke [1893, p. 381; see also Vogel 1892, p. 608]: "There remains, then, as the only intelligible rationale of solar sustentation, Helmholtz's shrinkage theory. And this has a very important bearing upon the nebular view of planetary formation; it may, in fact, be termed its complement." However, in the mid-nineteenth century the nebular hypothesis was still controversial and resisted by several leading astronomers and physicists. Thomson was one of the doubters.

Helmholtz returned to the Kant-Laplace hypothesis – "one of the happiest ideas in science" – in a public lecture he gave in Heidelberg in 1871. Apart from repeating his estimate of approximately 20 million years for the Sun's age, on this occasion he also looked into its future such as Thomson had done previously. Given that the mean density of the Sun was much smaller than that of the Earth, he considered the following scenario likely: "The sun will still continue in its condensation, even if it only attained the density of the earth … this would develop fresh quantities of heat, which would be sufficient to maintain for an additional 17,000,000 of years the same intensity of sunshine as that which is now the source of all terrestrial life" [Helmholtz 1995, p. 270]. All the same, in the end the Sun would become extinct and with it all life. "This is a thought which we only reluctantly admit," he added, for "it seems to be an insult to the beneficent Creative Power which we otherwise find at work in organisms and especially in living ones." But Helmholtz saw no way in which the human race could maintain its existence indefinitely.

---

[4] Helmholtz failed to distinguish properly between Kant's nebular cosmogony of 1755 and Laplace's hypothesis of 1796 of the formation of the planetary system. In reality the two theories have little in common, amongst other reasons because Kant suggested a speculative theory of the entire universe, and Laplace's more scientific theory was concerned only with the origin of the planetary system. On this question, see Stanley Jaki's introduction to [Kant 1981, pp. 13-76].

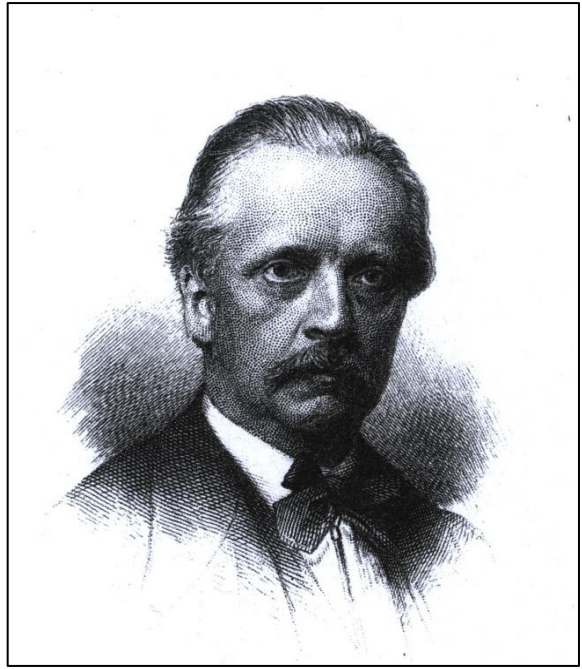

**Fig. 3**. Hermann von Helmholtz (1821-1894). From *Nature* **15** (1877), p. 389.

## 3.3   The later contraction theory

As we have seen, while Thomson had originally thought in terms of meteoric showers upon the Sun going on presently, by the 1860s he placed the hypothetical and very intense meteor showers in the past. At the beginning of the Sun's history it was heated in this way, and subsequently it had cooled and, as a result, contracted. On the assumption of constant solar density he found that "in contracting by one-tenth per cent in his diameter, or three-tenth per cent in his bulk, the sun must give out something either more, or not greatly less, than 20,000 years' heat" [Thomson 1891, p. 370]. Another of Thomson's (and Helmholtz's) assumptions was that the specific heat of the solar material was the same as that of water. Based on this simplifying assumption he deduced a thermal contraction of no less than one per cent per 860 years, which contradicted astronomical observations. For this and other reasons he adopted parts of Helmholtz's hypothesis of gravitational contraction.

   Thomson discussed other possibilities for the Sun's specific heat and also for the radial variation of its density, which was likely to increase toward the centre. He argued that there were strong reasons for believing that the specific heat varied between 10 and 10,000 times the value of water. Within this interval, he wrote in 1862, it "would follow with certainty that his [the Sun's] temperature sinks 100° Cent.

in some time from 700 years to 700,000 years." With regard to the chemical nature of the solar material, Thomson thought he had "excellent reason" for believing that it was similar to terrestrial matter. As reason for his confidence he referred to "the splendid researches of Bunsen and Kirchhoff," that is, to the very recent discoveries made with the spectroscope demonstrating the existence of iron, sodium, magnesium and other terrestrial elements in the Sun's atmosphere.

Contrary to Helmholtz's enthusiastic support of the nebular hypothesis, Thomson rather disliked it, stating in 1854 that it was "the reverse of the truth." His initial attitude to Helmholtz's solar theory was correspondingly negative; but, as mentioned, he gradually converted to a form of it. In an address to the Royal Institution of 1887, dealing with solar energy as "a splendid subject for contemplation and research in Natural Philosophy or physics," he even accepted the nebular hypothesis as "a necessary truth" [Thomson 1891, pp. 369-429]. He now refined his dynamical model of the Sun and its heat supply by taking into account that the mass density of the Sun was not uniform but increasing towards the centre. More importantly, he made use of a significantly larger value of the solar constant than the one obtained by Pouillet. From observations made on Mount Whitney the American astronomer Samuel P. Langley concluded in 1884 that the solar constant was 3.07 cal cm$^{-2}$ min$^{-1}$ or 1.7 times larger than Pouillet's value [Barr 1963]. (It later turned out that Langley's result was seriously wrong; the modern value of the "constant" is 1.361 kW m$^{-2}$ or 1.950 cal cm$^{-2}$ min$^{-1}$.)

Based on Langley's value and a more realistic assumption concerning the Sun's density Thomson reassessed the age of the Sun. He now calculated that the rate of gravitational contraction corresponded to an annual decrease of its radius of about 35 m, meaning that it "must have been greater by one per cent two hundred thousand years ago than at present." Thomson concluded that the Sun could not have existed as a strongly radiating body for more than 20 million years and that it would cease to shine appreciably in another 5 or 6 million years. Like Helmholtz, he was convinced that life on Earth, conditioned as it was upon an active Sun, would not continue indefinitely.

During the last decades of the nineteenth century the Helmholtz-Thomson contraction theory commanded a great deal of authority and was generally recognized as true by the majority of astronomers and physicists. Thomson's close friend and collaborator Peter Tait, professor of natural philosophy at the University of Edinburgh, fully supported the small time-scale of the Sun based on the contraction theory. In a lecture of 1874 he stated, without offering calculations, that it was "utterly impossible" that the Sun had heated the Earth for more than 15-20 million years [Tait 1876, p. 168]. Although several alternatives were proposed, these were either of an unappealing *ad hoc* nature or they were modifications of the contraction theory (see below). And yet the Helmholtz-Thomson theory was not entirely satisfactory. For one thing, it rested on a phenomenon – the gradual shrinking of the Sun – which lacked observational confirmation. In the 1870s the Italian astronomer and specialist in solar research, Pietro Angelo Secchi, thought to have detected variations in the solar diameter. However, a careful study made by the German astronomer Arthur Auwers, at the Berlin Academy of Science, failed to confirm the hypothesis that the Sun varies in size [Auwers 1873; Shahiv 2009, p. 39-41]. By the end of the century the consensus view was that the diameter of the Sun, as far as observations could tell, was constant.

Moreover, the contraction theory predicted a lifetime of the Sun much too limited to satisfy what most geologists and evolutionary biologists wanted, namely an age of the Earth of at least 100 million years. And yet, the contraction theory was sufficiently elastic that even the gap between 20 and 100 million years could be bridged. For example, if the Sun's intensity was assumed to have been smaller in the past, its lifetime could be considerably greater than the 20 million years. The American astronomer Charles Abbot, director of the Smithsonian Astrophysical Observatory, believed that the Sun had indeed been weaker in the past and that the standard theory could therefore be maintained.[5] Writing in 1911 he was aware of radioactivity as a possible source of the Sun's energy, but he considered it more satisfactory "to account for the solar heat by known causes rather than to invoke radio-activity of undiscovered materials." He consequently argued that "We may regard Helmholtz's contraction hypothesis as adequate to satisfy the requirements of geology and physics" [Abbot 1911, p. 278]. However, by that time a growing number of astronomers and physicists disagreed with Abbot's optimistic assessment of the Helmholtz-Thomson theory.

---

[5] Abbot's suggestion and the still earlier one of Perry (Section 4.1) were much later confirmed by Martin Schwarzschild in his classic *Structure and Evolution of the Stars* from 1958 where he concluded that the Sun's luminosity was considerably smaller in the past than its present value. Similar conclusions were independently arrived at by Fred Hoyle and Ernst Öpik and they eventually led to the so-called "faint young Sun paradox" which is still being discussed by astronomers and earth scientists [Feulner 2012].

## 3.4 The constitution of the Sun

In his essay of 1854 Helmholtz did not comment on the physical constitution of the Sun, whereas Thomson in 1862 assumed the solar matter to be incandescent and in a fluid state. The possibility of a gaseous Sun was suggested by the French astronomer Hervé Faye in 1865, but the model he presented was purely qualitative and conceptual [Clerke 1893, pp. 186-190]. The slow change from a liquid to a gaseous Sun during the last decades of the nineteenth century constituted an important advance in solar physics although it did not significantly affect the status of the contraction theory [see details in Robitaille 2011]. It was only with the work of the American scientist and engineer Jonathan Homer Lane that Clausius' mechanical theory of gases, first expounded in a paper of 1857, was applied to the Sun in the form of a mathematical model [Powell 1998]. "Some years ago," Lane [1870] wrote, "the question occurred to me in connection with this theory of Helmholtz whether the entire mass of the sun might not be a mixture of transparent gases." Not only did Lane discuss the distribution of density and temperature inside the Sun, he also took into account that the unknown solar gas might be either monoatomic or diatomic or a mixture of the two forms. It was known from Clausius' gas theory that the number of degrees of freedom $n$ depends on the ratio of the specific heats at constant pressure and constant volume, namely

$$\gamma = \frac{c_p}{c_v} = 1 + \frac{2}{n}$$

Thus $\gamma = 1.4$ for a diatomic gas and 1.67 for a monoatomic gas (until the discovery of argon in 1894 no gases of the latter class were known).

Unaware of Lane's earlier work the German physicist August Ritter wrote in 1878-1883 a series of papers in *Annalen der Physik* on stellar structure. Like Lane, he assumed the Sun to be gaseous and he gave a theoretical relation between its central temperature and the one at its surface. In a paper of 1883, which was later translated into English, Ritter suggested that the Sun should be considered as "a star gradually beginning to disappear." He shared with Thomson the belief in a future heat death. Ritter derived the time $T$ it takes for the Sun's radius to decrease from $r$ to its present radius $r_0$, which he took to be 688 million km. His result can be stated in the form

$$T = A\left[1 - \left(\frac{r_0}{r}\right)^{\alpha}\right]$$

Here *A* and *α* are quantities which depend on the atomicity of the solar gas. For $\gamma$ = 1.4 he found that for 5.5 million years ago the Sun's radius would have been about the same as the radius of the Earth's orbit ($r/r_0$ = 215); with Langley's new value of the solar constant instead of Pouillet's, the result became 3.67 million years. Although Ritter was aware that his theory did not allow him to calculate a maximum age of the Earth, he nonetheless found it "permissible to conclude from the above investigation that the actual age of the Earth must be far less than the estimates of some geologists, who place it at hundreds of millions of years" [Ritter 1898, p. 305]. That is, he joined Thomson in judging the discrepancy to be a problem for the geologists and not for the physicists. It should be noted that neither Lane nor Ritter expressed any doubts concerning the essential validity of the Helmholtz-Thomson theory of solar energy. They just took it for granted.

## 4  Voices of dissent

For a period of some forty years the Helmholtz-Thomson contraction theory was undoubtedly considered the standard theory of solar energy production. The two names associated with the theory, widely recognized as the greatest theoretical physicists of their time, added to its authority. But it was not the only theory of solar heat and especially in Great Britain there were many scientists who proposed alternatives to the gravitational contraction theory and its undesirable prophecy of an end to all life. Some of the dissenters wanted a mechanism for the Sun's energy that agreed with the evolution theory of the natural historians. Not all were taken seriously, but some, such as James Croll and William Siemens, were.

### 4.1  Newcomb and Perry

The distinguished American astronomer Simon Newcomb independently calculated from the contraction theory and the assumption of constant solar energy output that the *total* lifetime of the Sun was close to 18 million years. He estimated that the Sun could not have provided the Earth with approximately its present temperature for more than 10 million years [Newcomb 1878, pp. 505-511]. Newcomb, who at the time served as superintendent of the US Nautical Almanac Office, gave this figure in the widely read *Popular Astronomy*, a book first published in 1878 and subsequently in several revised editions. The same figure appeared in the 1892 edition of *Newcomb-Engelmann*, a detailed and highly recognized exposition originally based on Newcomb's book [Vogel 1892, p. 608]. Although Newcomb accepted the Helmholtz-Thomson solar theory and its basis in the nebular hypothesis, he cautiously pointed out that the latter hypothesis could not be regarded an established scientific fact. It was, he thought, rather "a result which science makes more or less probable, but of the validity of which opinions may reasonably differ."

What would happen with the heat of the Sun? Would it forever be lost? In discussing these questions Newcomb dismissed the idea that the dying Sun might receive new heat from some other celestial body, a star or a nebula, and also the idea that the Sun might be refuelled by "the ethereal medium." The latter hypothesis was possibly a reference to the Scottish engineer William Rankine, who in 1852, responding to Thomson's prediction of a heat death, speculated that "an opposite condition of the world may take place, in which the energy which is now being diffused may be re-concentrated into foci, and stores of chemical power again produced from the inert compounds which are now being continually formed." Rankine [1881, pp. 200-202] thought that such counter-entropic processes might take place by means of an ethereal medium filling the space between the stars. His hypothesis attracted critical interest. Helmholtz may have been aware of it, for in his 1854 lecture he referred to the ethereal medium transmitting rays of light and heat. "We know not," he said, "where the rays must return, or whether they eternally pursue their way through infinitude." But Clausius would have nothing of it. In a lengthy paper of 1864 he argued that if Rankine's hypothesis were correct "we would have to reject … the second law of the mechanical theory of heat" [Clausius 1864, p. 4; Kragh 2008, pp. 41-42].

Of more interest to Newcomb was the possibility that the radiant energy emitted by the Sun might return after a long cosmic journey: "Such a return can result only from space itself having such a curvature that what seems to us a straight line shall return into itself, as has been imagined by a great German mathematician." The mathematician he referred to was most likely Bernhard Riemann, who in an address of 1854 laid the foundation of non-Euclidean geometry as a mathematical discipline. Newcomb was one of the very few astronomers in the nineteenth century who took an interest in the new kind of geometry and its relevance for astronomy and physics [Kragh 2012]. However, he saw no good reason to accept that physical space was positively curved and hence finite. In fact, he dismissed the hypothesis of a cosmological return of solar energy as "too purely speculative to admit of discussion." In agreement with Thomson he concluded that "the radiant heat of the sun can never be returned to it," implying that the Sun had a finite and relatively short lifetime.

The discussion concerning Thomson's estimate of the Sun's age was in many ways secondary to that of the Earth's age. On the other hand, the first age evidently put limits to the second age. The engineer and physicist John Perry, who had once served as Thomson's assistant, criticized the simplifying assumptions that led to the young Earth, suggesting that if the assumptions were changed the age of the Earth might become several hundreds of million years. "The argument from the sun's heat seems to me quite weak," Perry wrote in a letter to Tait of 1894. "Even a geologist without mathematics can see that the time given by Lord Kelvin will be increased if we assume that in past times the sun radiated energy at a smaller rate than at present … and the rate may have greatly varied from time to time." But Thomson, who by then had become Lord Kelvin, had made up his mind. In a letter to Perry from the same year, he insisted that his calculations based on the solar contraction theory were correct: "Helmholtz, Newcomb, and another [Thomson himself], are inexorable in refusing sunlight for more than a score or a very few scores of million years of past time" [Perry 1895; Burchfield 1975, pp. 134-139; Smith and Wise 1989, pp. 544-548].

## 4.2 Siemens' alternative

The scattered opposition against the Helmholtz-Thomson theory of solar energy came primarily from amateur scientists and people outside the communities of physics and astronomy. One of them was William Mattieu Williams, an author and amateur chemist, who in 1870, in a book titled *The Fuel of the Sun*, developed a speculative theory that allowed the Sun to radiate indefinitely at its present rate. Williams' solution was to postulate a universal atmospheric medium of very low density which received the Sun's heat and continually reemitted it to the Sun, making the solar system a veritable *perpetuum mobile*. Although speculative and amateurish, by the standards of the time Williams' book [1870] qualified as science. It was accordingly reviewed in *Nature* by the respected physicist Balfour Stewart, who with an understatement judged that Williams' hypothesis "is [not] an improvement upon that of Helmholtz and Thomson." Another amateur solar scientist, but of a calibre and status quite different from the obscure Williams, was the German-born engineer, inventor and industrialist William Siemens (a brother of Werner von

Siemens), who had revolutionized the steel-manufacturing process with the invention of the regenerative furnace.

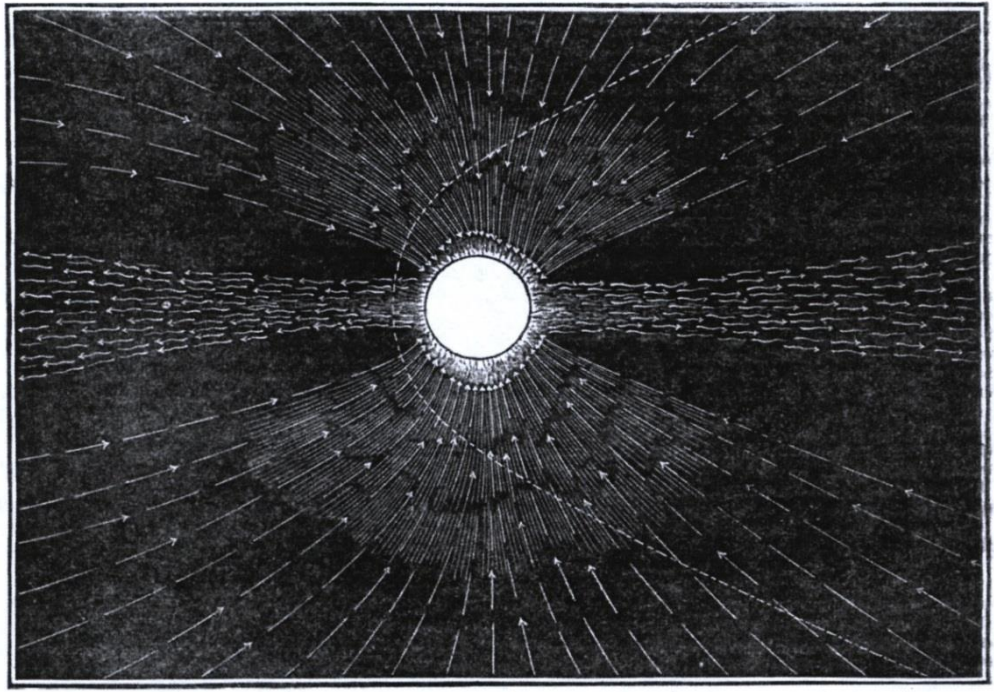

**Fig. 4**.  William Siemens' model of the regeneration of solar energy.
From *Nature* **25** (1882), p. 441.

Trained as a mechanical engineer, Siemens admitted that he could not "claim intimate acquaintance with the intricacies of solar physics." Nonetheless, he was a respected member of the British scientific community and since 1862 a Fellow of the Royal Society. Siemens noted that less than a millionth part of the Sun's energy was intercepted by the planets, while the rest was radiated into space and apparently lost to the solar system. A cosmic machine of such poor efficiency and so little practical purpose offended his engineering mind, and he consequently devised a more efficient model of the Sun. He was convinced "that the prodigious and seemingly wanton dissipation of solar heat is unnecessary to satisfy accepted principles regarding the conservation of energy, but that this heat apparently expended without producing any effect whatever may be effectively arrested and return over and over again to the sun, in a manner somewhat analogous to the action of the heat recuperator in the regenerative gas furnace" [Siemens 1882, p. 397; Schaffer 1995]. Siemens' paper was considered important enough to be translated into French in *Annales de Chimie et de Physique*.

According to Siemens, space was filled with a highly rarefied gas mixture of density equal to or less than $5 \times 10^{-4}$ times that of the Earth's atmosphere; he stated the lower limit to be about 100 times less. The gases were gravitationally drawn into the Sun at its poles, where they would decompose under the high temperature, and as the result of the Sun's rotation the combustion products would be ejected at the solar equator and redistributed through space. "What would become of these products of combustion when thus rendered back into space?" Siemens asked. His answer: "The solar radiation would … step in to bring back the combined materials to a condition of separation by a process of dissociation carried into effect at the expense of that solar energy which is now supposed to be lost to our planetary system." The model was as ingenious as it was artificial. It got rid of the Sun's distasteful wastefulness and kept the solar system going nearly indefinitely, the only lost part of the radiation being that which was not absorbed by the cosmic medium. But this was only possible at the expense of introducing arbitrary hypotheses and violating the second law of thermodynamics.

Nonetheless, Siemens' cyclical-chemical theory aroused much interest and a few scientists supported its basic features. Among them was Thomas Sterry Hunt, an American chemist and geologist with an inclination for speculative theories. Most leading physicists and astronomers either ignored Siemens' solar theory or they politely dismissed it as wrong and based on false premises. Hervé Faye, professor at the École Polytechnique in Paris and an astronomer of international repute, objected that even a gaseous medium as tenuous as the one proposed by Siemens would disagree with the observed behaviour of Encke's comet and have other drastic astronomical consequences. Moreover, Siemens had neglected to take into account the enormous quantity of matter that his hypothesis would add to the Sun and other stars. "It is not very probable that the astronomers will adopt such hypotheses," Faye said with an understatement. "No doubt they would be pleased to think that Nature has provided the Sun with resources to make his heat last longer; but as his final refrigeration is still, under any circumstances, a tolerably distant catastrophe, they will console themselves by the thought that the things of this world, even the most beautiful, do not appear to be made for ever."[6]

---

[6] Communication in *Comptes Rendus* of 9 October 1882. Together with other sources on Siemens' solar theory it was translated in Siemens [1883]. William Siemens died in 1883 and his theory essentially with him.

Charles Young, professor of astronomy at Princeton University, was highly regarded as a specialist in solar physics. In a monograph of the Sun first published in 1881 he presented the contraction theory as far the most convincing explanation of the Sun's energy if not necessarily the correct explanation. He took Siemens' theory seriously as a possible alternative and dealt with it at length, but only to raise objections against it similar to those by Faye. "One almost regrets that the theory can not be accepted," Young [1896, p. 322] wrote.

**4.3 Other alternatives: Croll and Chamberlin**

Best known for his astronomical theory of terrestrial climate change, including the ice ages, the self-educated Scottish geologist James Croll offered another and less unorthodox alternative to the Helmholtz-Thomson standard theory. But his aim was the same, to suggest a theory that could account for almost any duration of solar heat. Croll [1878] admitted that "Gravitation is now almost universally appealed to as the only conceivable source from which the sun could have obtained his energy." He objected that, because the physicists could not conceive of any other source than gravitational contraction, it did not follow there is no other source. Referring to the demands of the geologists and natural historians, he wrote: "If 20,000,000 or 30,000,000 years do not suffice for the evolution theory, then either that or the gravitation theory of the origin of the sun's heat will have to be abandoned." In a later paper he developed his impact theory into a cosmological scenario which "on purely scientific grounds" led to an absolute beginning of the evolutionary universe. According to Croll [1889], it followed that the universe would continue its evolution forever, whereas the thermal equilibrium state known as the heat death would never occur.

Croll, who evidently considered the biological evolution theory to be better founded than the Helmholtz-Thomson solar theory and the nebular hypothesis, proposed that the energy source for the Sun's heat was kinetic rather than based in gravitational potential energy. Moreover, he thought it was instantaneous rather than acting continually. According to his collision theory, the Sun's heat was originally derived from a nebulous mass made incandescent by the collision of two bodies moving in space at very great speed. For two colliding bodies, each having half the Sun's mass, he calculated that velocities of 476 miles per second would result in an amount of heat allowing the Sun to radiate at its present rate for 50 million years. And he could easily imagine greater speeds and hence a greater age. "If we assume the original velocity to have been 1,700 miles per second, an amount of heat would be generated in a single moment which would suffice for no less than 800,000,000 years" [Croll 1878].

What mattered to Croll was that his collision theory "was more in harmony with the principles of evolution than the gravitational theory," but this was an argument that did not impress the physicists. As they saw it, there were obvious difficulties with Croll's alternative. For one thing, from where did the colliding bodies originate and how did they acquire their excessive velocities? For another thing, the theory was purely qualitative and hardly testable. It could result in almost any duration of the Sun's heat, depending on the arbitrarily chosen and suspiciously high original velocities. Celestial bodies with velocities of this magnitude were unknown to the astronomers. Thomson, who preferred to ignore Croll's challenge, only referred to the collision theory in 1887 and then without mentioning Croll by name. He [1891, p. 405] concluded that the high-speed collision between two primeval stellar bodies was "exceedingly improbable."

The American geologist Thomas C. Chamberlin, professor at the University of Chicago and a specialist in glacial geology, was a major player in the discussion of the origin of the Earth and its relation to the nebular hypothesis [Winnek 1970; Brush 1996, vol. 3]. By the end of the nineteenth century he concluded from geological evidence that the nebular hypothesis – "this theory of a simple decline from a fiery origin to a frigid end" – was seriously wrong, and he thought that the same was the case with Thomson's theory of the Earth and its age. In a paper of 1899 he attacked not only the orthodox theory of an originally molten Earth but also the associated Helmholtz-Thomson theory of the Sun's energy and age. Chamberlin's critique was essentially negative, as it consisted in questioning the completeness of the solar theory and the unspoken assumptions upon which it rested. He argued, as Perry and others had done previously, that the unknown distribution of solar heat in time was essential to how the age of the Sun could serve as a "moral guiding force" to the history of the Earth. Chamberlin did not come up with an alternative solar theory except that he vaguely speculated that somehow the Sun's energy might be reflected back to the solar system by some particular configuration of the nebulae. His geological and cosmological ideas were influenced by his religious belief, which can be summarized as a mixture of pantheism and Arminian Christianity [Winnek 1970]. For moral and religious reasons he thought that the evolution of nature was tightly connected with the existence of humans, and he consequently denied the cosmic heat death caused by the second law of thermodynamics.

Given the almost complete lack of knowledge of the state of matter in the interior of the Sun, how could Thomson speak with such an air of certainty about the origin of solar heat? "What the internal constitution of the atoms may be is yet an open question," Chamberlin [1899, p. 12] pointed out. He elaborated:

> It is not improbable that they are complex organizations and the seats of enormous energies. Certainly, no competent chemist would affirm either that the atoms are really elementary or that there may not be locked up in them energies of the first order of magnitude. … Nor would he probably be prepared to affirm or deny that the extraordinary conditions which reside at the center of the sun may not set free a portion of this energy. … Why should not atoms, atomecules [sic], and whatever lies below, one after another have their energies squeezed out of them; and the outer regions be heated and lighted for an unknowable period at their expense?

Although Chamberlin's comments were nothing but speculations, and not even very original speculations, they turned out to be prophetic. The term "atomecule" was invented by Chamberlin, perhaps in the meaning of a hypothetical subatomic entity. Croll [1889, pp. 71-110] discussed at length various hypotheses of the complex atom and what he called "the pre-nebular condition of matter." At the time it was often assumed that the evolution of chemical elements was a natural complement to the nebular hypothesis and even to Darwinian evolution. "Inorganic Darwinism" was in vogue.

At the time Chamberlin launched his attack against Thomson he was about to develop a strong alternative to the nebular hypothesis of the formation of the Earth. Together with his young Chicago colleague, the astronomer Forest Ray Moulton, he proposed a "planetesimal theory" according to which the planets were formed by aggregation of small solid particles and not by condensation of an original gaseous material. For a period the Chamberlin-Moulton theory was highly regarded not only by geologists but also by American astronomers. However, the theory was concerned only with the formation of planets and not with the source of solar energy. On the other hand, in a textbook of celestial mechanics Moulton [1914, p. 66] made it clear that he had no confidence in the Helmholtz-Thomsen theory but preferred an explanation of the Sun's heat in terms of radioactivity and subatomic energy.

## 5  Some wider issues

As we have seen, during the second half of the nineteenth century physicists and astronomers reached the conclusion that the Sun was doomed to die and all life come to an end. The scenario was often, if far from always, seen as a consequence of the general heat death of the universe derived from the authoritative second law of thermodynamics. The heat death scenario was widely discussed not only by scientists but also by philosophers, authors, theologians, and social critics. Balfour Stewart [1870, p. 271], professor of natural philosophy in Manchester and a collaborator of Tait, formulated the hypothesis as follows:

> Will the sun last for ever, or will he also die out? There is no apparent reason why the sun should form an exception to the fate of all fires, the only difference being one of size and time. It is larger and hotter, and will last longer than the lamp of an hour, but it is nevertheless a lamp. The principle of degradation would appear to hold throughout, and if we regard not mere matter but useful energy, we are driven to contemplate the death of the universe.

The extinction of life caused by the Sun running out of fuel was widely considered unacceptable for emotional or ideological reasons and that even though the ultimate catastrophe was millions of years in the future. Numerous proposals were made to avoid the heat death or, more specifically, the death of the Sun. Rankine's ether-speculation of 1852 was the first, and it was followed by dozens of other proposals. The Austrian physical chemist Joseph Loschmidt suggested that the Sun and the other stars would evolve cyclically over long periods of time, never to become permanently extinct. They would continually change between being hot and cold, and bright and dark, a speculation he found evidence for in the existence of novae and so-called dark stars. In this way Loschmidt believed to have destroyed "the terrifying nimbus of the second law." Humankind, he wrote in 1876, "could take comfort in the disclosure that humanity was not solely dependent upon coal or the Sun in order to transform heat into work, but would have an inexhaustible supply of transformable heat at hand in all ages" [Daub 1970, p. 221].

Some of the mechanisms invoked to avoid the heat death relied on the idea that energy would be rejuvenated by collisions between stars. Mayer, the discoverer of the principle of energy conservation, preferred a universe in a steady state and consequently rejected the heat death of Clausius and Thomson. In a letter of 1869 he wrote to a friend that "a few years ago Brayley in London said that if two stars of the size of the Sun collide, all condensed mass must be dissolved and the molecules be dissipated throughout the universe" [Mayer 1893, p. 301]. The reference was to Edward William Brayley, a London professor of physical geography according to whom the planets were formed by collision and coalescence of meteors originally produced by solar matter. Brayley [1865] speculated that the Sun consisted of a highly rarefied imponderable substance – "still more transcendently and intensely solid and elastic" than the ordinary ether – out of which ponderable matter was generated. Mayer thought that a kind of collision hypothesis, another version of which was advocated by Croll, made the heat death avoidable.

To move from the scientists to the philosophers, at about the same time the influential British philosopher Herbert Spencer took notice of the new developments in physics and astronomy. As he commented in one of his major works, the "tacit assumption hitherto current, that the Sun can continue to give off an undiminished amount of light and heat through all future time, is fast being abandoned." Spencer was a liberal thinker who firmly believed in progressive evolution in both nature and society. He was fascinated by the principle of energy conservation but unpleasantly surprised when he realized that the other great law of thermodynamics was probably incompatible with continual evolution and progress. Spencer [1867, p. 493 and p. 514] described his uneasiness as follows:

> If the solar system is slowly dissipating its forces – if the Sun is losing his heat at a rate which will tell in millions of years – if with diminution of the Sun's radiation there must go on a diminution in the activity of geologic and meteorologic processes as well as in the quantity of vegetal and animal existence – … are we not manifestly progressing towards omnipresent death?

The affirmative answer following from the second law of thermodynamics was unacceptable to Spencer. Unable to make the second law comply with his belief in progressive evolution, he concluded that the heat death was wrong for philosophical reasons. The Sun *had* to shine for ever. His anxiety with regard to the cosmic future was shared by Charles Darwin, who agreed with Spencer that the evolutionary process would not only lead to greater complexity but also to greater human perfection. He found the prospect of an end to life to be an "intolerable thought." In his posthumously published autobiography begun in 1876 Darwin referred to "the view now held by most physicists, namely, that the sun with all the planets will in time grow too cold for life, unless indeed some great body dashes into the sun and thus gives it fresh life."[7] The father of modern evolutionary biology found the physicists' view to be horrifying. In a letter to the evolutionist Alfred Wallace of 1869 he admitted that "Thomson's views of the recent age of the world have been for some time one of my sorest troubles."

As a final example of the cultural pervasiveness of solar physics in the period, consider Friedrich Engels, Karl Marx's close collaborator and a leader of the early Communist movement [Foster and Burkett 2008]. In a letter to Marx of 1869 he expressed his dislike of the idea of steadily increasing entropy and its long-term cosmological consequences. He referred to Clausius' heat death, which he described as "a very absurd theory, which incidentally follows with a certain inevitability from Laplace's old hypothesis."[8] Engels thought that the law of entropy increase was incompatible with dialectical materialism and ideologically dangerous because of its association with creation, miracles and theism. In his fragmentary notes posthumously published as *Dialectics of Nature* he speculated that the solar heat radiated into space "must be able to become transformed into another form of motion, in which it can once more be stored up and rendered active." As another mechanism he suggested that dead stars would sooner or later collide with one another and produce an enormous heat energy that locally would lower the entropy and restart evolutionary processes. The life of the Sun would come to an end, and the same would happen for the other stars, yet the infinite universe would remain alive for ever. It was not the first time that such a scenario was proposed, and it was not the last.

# 6   The radioactive Sun

"The difficulty about the Sustenance of the Sun's heat has been removed … There is no doubt whatever that the theory of Helmholtz gives the true explanation of the great problem of the source of the Sun-heat." This is how the Irish astronomer Robert Stawell Ball [1893, pp. 270], newly appointed Professor of Astronomy and Geometry at Cambridge University, confidently described the Helmholtz-Thomson contraction theory in a popular book on the Sun. As to the original source of the Sun's energy, he suggested that it was due to "a collision which occurred many millions of years ago between two dark bodies which … happened to approach sufficiently close together." Ten years later Ball was not so sure. He now suggested that a better candidate for the Sun's production of energy might be found in the new and sensational phenomenon of radioactivity. In a letter of 1903 to his friend and compatriot John Joly, a physicist and geologist at Trinity College, Dublin, he wrote [Hufbauer 1981, p. 281]:

> Have you seen radium? It certainly gets over the greatest of scientific difficulties, viz. the question of sunheat. The sun's heat cannot have lasted over 20,000,000 of years if it is due to contraction. But the geologists would have, say, 200,000,000. Now the discrepancy vanishes if the sun consists in any considerable part of radium, or something that possesses the like properties. It is a most instructive discovery.

---

[7] Darwin's autobiography and other works and sources by Darwin or related to him can be found online: http://darwin-online.org.uk/ (accessed July 2016).

[8] Quoted from http://www.marxists.org/archive/marx/index.htm (accessed July 2016). This source also includes an English translation of Engels' *Dialectics of Nature*.

We witness a similar change of mind in the famous writer H. G. Wells, who in 1902 gave a lecture to the Royal Institution on "The Discovery of the Future." He argued that "there is reasonable certainty that this sun of ours must radiate itself toward extinction and that this earth of ours … will be dead and frozen, and all that has lived upon it will be frozen and done with." Eleven years later, when the lecture was reprinted, Wells added a footnote: "Not now, … the discovery of radio-activity has changed all this" [Mousoutzanis 2014, p. 84]. Incidentally, in his novel *The World Set Free* of 1914, Wells spoke of "atomic bombs" and the possibility of "tapping the internal energy of atoms."

## 6.1  Early research in radioactivity

It took a couple of years until radioactivity – originally known as "uranium rays" or "Becquerel rays" – attracted wide interest among physicists and chemists. It was only with Marie and Pierre Curie's discovery in 1898 of two hitherto unknown elements much more active than uranium (radium and polonium), that radioactivity made headlines and was eagerly investigated by a large number of researchers. By the early years of the new century Ernest Rutherford and his collaborator Frederic Soddy had established that radioactive substances decay spontaneously and apparently randomly. The atoms strangely transmuted into other atoms, with nature performing the alchemist's art. The phenomenon became even stranger when Marie Curie and her assistant Albert Laborde reported in 1903 that the heat energy liberated by radium was enormous. Using an ice calorimeter they found that radium (together with its decay products) generated about 100 calories per gram per hour, or about 200,000 times more than the total burning of coal.

Radioactivity was of great interest not only to physicists and chemists but also to geologists, meteorologists and astronomers. For one or two decades it was widely believed to have a cosmic significance that included the solar system and even the stellar universe as a whole [Hufbauer 1981; Kragh 1997]. The belief rested on two widely accepted but wrong assumptions. First, it was generally assumed that the abundance of the elements in the Earth's crust was largely the same as in the solar system. Although Rutherford [1962, p. 784] admitted that there was no "direct evidence that radioactive matter exists in the sun," in 1905 he argued that, "from the similarity of the chemical constitution of the sun and the earth, its presence is to be expected." He stated that an amount of 2.5 ppm of radium in the Sun would account for its present rate of energy emission. The eminent astronomer and geophysicist George Howard Darwin, a son of the famous naturalist, had since the mid-1880s been critically interested in the contraction theory and he now glimpsed an alternative to it. In his presidential address of 1905 to the British Association, he confirmed the hypothesis of radioactive stars: "Now we note that the earth contains radio-active materials, and it is safe to assume that it forms in some degree a sample of the materials of the solar system. Hence it is almost certain that the sun is radioactive also." Darwin [1905, p. 29] further argued that probably the abundance of radioactive substances on the Sun was greater than on the Earth.

Second, many scientists believed that radioactivity was a property common to all matter, only appearing in different degrees of intensity. From an extensive series of experiments the Cambridge physicist Norman Campbell concluded in 1906 that most likely ordinary matter was weakly radioactive. Rutherford [1906, pp. 217-218], still at McGill University in Montreal, put his authority behind the claim: "The evidence obtained by him [Campbell] affords very strong proof that ordinary matter does possess the property of emitting ionizing radiations, and that each element emits radiations differing both in character and intensity." If all matter were radioactive, ordinary substances would have a finite lifetime, which Rutherford estimated to "at least one thousand times that of uranium, i.e. not less than $10^{12}$ years." The question of whether ordinary matter is radioactive or not remained unsettled for some years, but about 1915 the consensus view was that the large majority of chemical elements were non-radioactive.

A more specific reason for assuming the Sun to be a radioactive body was the insight that helium, the solar element discovered in 1895, was liberated from radium and other radioactive bodies in the form of alpha rays. Evidence for helium as a disintegration product of radioactive change was first established by Soddy and William Ramsay in 1903, but it took a few more years until the identification of alpha particles and helium ions ($\alpha = He^{2+}$) was definitely confirmed. It was tempting to infer that the Sun's content of helium derived from radioactive sources hidden in its interior. This is what the American physical chemist Harry Jones [1903, p. 281] did. "If all the helium in the sun comes from radium," he wrote, "then there must be, or at least must have been, enormous quantities of radium in the sun." He suggested that radium decay might be a better mechanism for the Sun's heat than Helmholtz's contraction mechanism.

Given the short lifetime of radium and its apparent absence from solar spectra the Sun could not be powered by radium alone, but there were other possibilities for a radioactive Sun. As Rutherford and a few other scientists speculated, perhaps ordinary elements in the intensely hot Sun broke down to alpha-active elements of a nature similar to radium but with much longer lifetimes. This is what he suggested in his 1913 monograph *Radio-Active Substances and Their Radiations*: "If the atomic energy of the atoms is available, the time during which the sun may continue to emit heat at the present rate may be much longer than the value computed from ordinary dynamical data" [p. 656].

**6.2   The Sun as a radioactive machine**

From about 1902 to 1914 the cosmic role of radioactivity was widely discussed by physicists and astronomers in connection with the problem of stellar energy. The letter section of the journal *Nature* of 1903 contained several communications on the issue, including arguments why the rays from radioactive bodies supposedly emitted by the Sun had not been detected at the surface of the Earth. Joly [1903] summarized the case for radium's presence in the Sun as follows: "(1) The presence of radium on the earth; (2) the high atomic weight of radium; (3) Arrhenius' theory of the Aurora Borealis; (4) the fact that the estimate of the duration of solar heat from the dynamical source appears to run counter to geological data." To Joly's list might be added that a radioactive Sun, combined with the disintegration hypothesis and the composite atom, promised an understanding of how the chemical elements had come into being. To the German physicist Johannes Stark, a future Nobel Prize laureate of 1919, stellar energy production was closely related to element synthesis. He speculated [Stark 1903]:

> As the transformation of atoms in some elements is still going on, it may be supposed that there was a time when our chemical atoms did not exist in the present amount, while other types of matter were more common. In the later change of the arrangement of the positive and negative electrons, or in the genesis of the present chemical atoms, a very large amount of the potential energy of their electrons was transformed to kinetic energy. … It is reasonable to suppose that the temperature of the Sun and stars is partly due to the genesis of chemical atoms.

The idea of atoms made up of negative and positive electrons was entertained by several physicists in the early twentieth century, among them James Jeans. At the time a "positive electron" might either refer to a hypothetical mirror particle of the negative electron (a positron *avant la lettre*) or to the much heavier hydrogen ion $H^+$, or what was later called the proton.

Soddy was among those who thought that the stars were powered by radioactive decay processes. "It seems justifiable," he wrote in a progress report on radioactivity from 1905, "to assume that in the stars, as in the earth, evolution is proceeding from the heavy to the lighter forms" [Soddy 1975, on p. 87]. For a brief period of time the fascinating phenomenon of radioactivity was discussed enthusiastically as if it provided new answers to nearly all problems of science. One of them was the age of the Sun and its source of energy. In his presidential address to the British Association in 1906, the distinguished zoologist Ray Lankester joined the chorus of radioactive enthusiasts. "If the sun consists of a fraction of one per cent of radium, this will account for and make good the heat that is annually lost by it," he claimed. Lankester [1906, p. 325] was pleased to point out how radioactivity solved the controversy over the age of the Earth and vindicated the view of the natural historians:

> This is a tremendous fact, upsetting the calculations of the physicists as to the duration in past and future of the sun's heat and the temperature of the earth's surface. The physicists … have assumed that its material is self-cooling … [but] it has now, within these last five years, become evident that the earth's material is *not* self-cooling, but on the contrary self-heating. And away go the restrictions imposed by physicists on

> geological time. They are now willing to give us not merely a thousand million years, but as many as we want.

The aging Thomson denied that radioactivity forced a reconsideration of the validity of the contraction theory. In a letter of 1906 he made his position clear: "The gravitational theory is amply sufficient to account for the heat of both bodies [Sun and Earth], and all of the stars in the Universe, and it seems almost infinitely improbable that Radium adds practically to their energy for emission of heat and light" [Smith and Wise 1989, p. 550]. On the other hand, the great French mathematician and theoretical physicist Henri Poincaré [1911, pp. 190-220] found the suggestion of a radioactively fuelled Sun to be attractive. Admitting that it was hypothetical and perhaps premature, he nonetheless considered the radioactive hypothesis to be a far better candidate for the Sun's energy production than the Helmholtz-Thomson theory. However, at the time Poincaré expressed his cautious support for the radioactive Sun, the hypothesis was slowly in decline.

It was a problem for the hypothesis that spectral lines from known radioactive elements had not been detected from either the Sun or other stars. But in 1912 a German astronomer at the Bonn Observatory, Hermann Giebeler, reported that he had observed lines in the spectrum of the recently discovered Nova Geminorum 2 which revealed the presence of uranium, radium and radon (at the time called radium emanation). In the same year the German physicist and authority in spectroscopy Heinrich Kayser indirectly supported Giebeler's evidence by suggesting a theory of the origin of novae based on radioactive processes [Giebeler 1912; Kayser 1912]. However, the evidence was disputed by other astronomers and soon considered to be erroneous. The lines reported by Giebeler could not be found at either the Mount Wilson Observatory [Adams and Kohlschütter 1912] or the Yerkes Observatory. As Samuel Mitchell [1912] at the latter observatory pointed out, "the apparent coincidences between the wave-lengths of two spectra have many times in the history of spectroscopy led to false identifications." He also used the occasion to dispute an earlier claim by the British astronomer Frank Dyson of radium lines in the Sun's chromosphere. The following year Mitchell [1913] summarized: "From theoretical considerations we are positively convinced that there must be radium in the sun. But to prove this is another problem!"

## 6.3 An unburied corpse

Rutherford [1962, p. 785] realized that with the new energy source the heat of the Sun would endure for a much longer time than previously thought, but eventually it would come to an end: "Science offers no escape from the conclusion of Kelvin and Helmholtz that the sun must ultimately grow cold and this earth must become a dead planet moving through the intense cold of empty space." Whereas some scientists interpreted the irreversibility of radioactive processes as evidence against an eternal universe, others speculated that new and purely hypothetical forms of radioactivity might justify an eternally cyclic universe. In *The Interpretation of Radium*, a book based on a series of public lectures given at the University of Glasgow in 1908, Soddy [1909, pp. 241-242] wrote:

> The idea which arises in one's mind as the most attractive and consistent explanation of the universe in light of present knowledge, is perhaps that matter is breaking down and its energy being evolved and degraded in one part of a cycle of evolution, and in another part still unknown to us, the matter is being built up with the utilisation of waste energy. The consequences would be that, in spite of the incessant changes, an equilibrium condition would result, and continue indefinitely.

Imagining a universe operating by means of atom-destructive and complementary atom-constructive processes, both conceived as radioactive, Soddy thought that in this way he had conceived a universe "demanding neither an initial creative act to start it nor a final state of exhaustion as its necessary termination." Another future Nobel Prize laureate in chemistry, Walther Nernst, entertained somewhat similar ideas.

The radioactive Sun, and generally the possibility that stellar energy might be due to spontaneous subatomic processes, attracted much attention as an alternative to the Helmholtz-Thomson contraction theory. Based on Langley's solar constant and new assumptions concerning the density of the inner part of the Sun, Darwin [1903] calculated that the contraction theory allowed the Sun to shine at its present rate for only 12 million years. He consequently appealed to subatomic energy as a solution to the problem, but his appeal was characteristically vague: "I think we have no right to assume that the sun is incapable of liberating atomic energy to a degree at least comparable with that which it would do if made of radium. Accordingly, I see no reason for doubting the possibility of augmenting the estimate of solar heat as derived from the theory of gravitation by some such factor as ten or twenty." Interest

39in the radioactive Sun was short-lived and often limited to a qualitative and rhetorical level. The hypothesis lacked observational support and was never developed into a proper theory. Although by 1915 it was effectively abandoned and soon forgotten, in a sense it can be considered a precursor to the new ideas of solar energy based on subatomic processes that were pioneered in the early 1920s.

This new chapter is well described in the literature [Hufbauer 2006; Wesemael 2009] and I shall not enter it except pointing out the crucial role played by Eddington [1917; 1920]. As early as 1917 he suggested that electron-proton annihilation might provide the "almost inexhaustible store of energy" that was so obviously lacking in the contraction theory. An age of the Sun of only 20 million years could no longer be taken seriously. Or, as he phrased it in an important address given to the British Association in 1920, "Lord Kelvin's date of the creation of the sun is treated with no more respect than Archbishop Ussher's." Eddington described the time-honoured theory of Helmholtz and Thomson as follows:

> If the contraction theory was proposed to-day as a novel hypothesis I do not think it would stand the smallest chance of acceptance. … So far as it is possible to interpret observational evidence confidently, the theory would be held to be negatived definitely. Only the inertia of tradition keeps the contraction theory alive – or, rather, not alive, but an unburied corpse.

## 7  Conclusions

The problem of the Sun's energy production was essentially solved in 1939 almost a century after Mayer had first addressed it on the basis of the law of energy conservation. From about 1850 to the early twentieth century the favoured explanation was that solar thermal energy arose from a decrease in gravitational potential energy, which for most of the period was ascribed to a slight contraction of the Sun. Despite several alternatives, some of them of a decidedly speculative nature, the Helmholtz-Thomson contraction theory (which had roots back to Kant) enjoyed an authoritative status for about forty years. The theory was based on well-established mechanical physics, was elegant and conceptually simple, and had a great deal of explanatory power. Moreover, it fitted well with the popular nebular world view and generally with the *Zeitgeist* characterizing the Victorian era. The problem of the Sun's heat attracted intense interest far outside the small communities of physicists and astronomers.

Despite its scientific authority the gravitational contraction theory was not directly verifiable and largely unsuccessful as far as testable predictions were concerned. First of all, the theory led to an age of the Sun, and consequently also for the Earth, that at the turn of the century was realized to be much too small. Nevertheless, it was not abandoned but continued to live on for some time if only as a shadow of its former glory. The major reason was that there was no alternative theory to replace it. Rather than being falsified, physicists and astronomers lost confidence in it, instead speculating that some sort of unspecified subatomic energy might explain the Sun's prodigious output of energy. Only with the *annus mirabilis* of nuclear physics in 1932 could these speculations be developed to a proper theory of stellar energy based on thermonuclear reactions.

The Helmholtz-Thomson contraction theory is usually and with good reason considered one the many mistakes of which the history of science is so rich. But it may be argued that it was only a mistake in so far that it was thought to apply to the Sun. In a certain sense the Victorian contraction theory was correct and it has survived in modern astrophysics under the somewhat unfortunate name the Kelvin-Helmholtz (KH) mechanism. The corresponding KH time scale refers to the time it takes for a stellar body to radiate away at a constant rate all of its potential energy. With *L* denoting the average luminosity the time scale is given by

$$\tau_{\text{KH}} = \frac{GM^2}{RL}$$

If one imagines the Sun to cease burning nuclear fuel it would contract and continue to radiate its store of gravitationally bound energy for approximately 10 million years. The Helmholtz-Thomson contraction theory applies to pre-main sequence stars where the energy source is gravitational contraction until they grow hot enough to start thermonuclear reactions; it applies to brown stars; and it also applies to gas giant planets such as Jupiter, causing it to shrink at a rate of 1 mm per year [Irwin 2009, p. 4].

**References**
Abbot, C. G. 1911a. The solar constant of radiation, *Proc. Amer. Phil. Soc.* **50**: 235-245.
Abbot, C. G. 1911b. *The Sun*, Appleton and Co., New York.

Adams, W. S. and A. Kohlschütter 1912. Observations of the spectrum of Nova Geminorum no. 2, *Astrophys. J.* **36**: 293-321.

Armstrong, W. G. 1864. [Presidential address], *Report, Brit. Assoc. Adv. Sci.* **34**: li-lxiv.

Auwers, A. 1873. On the alleged variability of the Sun's diameter, *Month. Not. Roy. Astron. Soc.* **34**: 22-24.

Ball, R. S. 1893. *The Story of the Sun*, Cassell and Company, London.

Barr, E. S. 1963. The infrared pioneers, III: Samuel Pierpont Langley, *Infrared Phys.* **3**: 195-206.

Becker, G. F. 1898. Kant as a natural philosopher, *Amer. J. Sci.* **5**: 97-112.

Brayley, E. W. 1865. Inferences and suggestions in cosmical and geological philosophy, *Proc. Roy. Soc.* **14**: 120-129.

Brush, S. G. 1987. The nebular hypothesis and the evolutionary worldview, *Hist. Sci.* **25**: 245-278.

Brush, S. G. (1996). *A History of Modern Planetary Physics*, vols. 1-3, Cambridge University Press, Cambridge.

Burchfield, J. D. 1975. *Lord Kelvin and the Age of the Earth*, University of Chicago Press, Chicago.

Caneva, K. L. 1993. *Robert Mayer and the Conservation of Energy*, Princeton University Press, Princeton.

Cantor, G. 1983. *Optics after Newton: Theories of Light in Britain and Ireland, 1704-1840*, Manchester University Press, Manchester.

Chamberlin, T. C. 1899. Lord Kelvin's address on the age of the Earth as an abode fitted for life, *Science* **9**: 889-901, **10**: 11-18.

Clausius, R. 1864. Ueber die Concentration von Wärme- und Lichtstrahlen und die Gränzen ihrer Wirkung, *Ann. Physik und Chemie* **121**: 1-44.

Clerke, M. A. 1893. *A Popular History of Astronomy during the Nineteenth Century*, Adam & Charles Black, Edinburgh.

Croll, J. 1878. Age of the Sun in relation to evolution, *Nature* **10**: 206-207, 321, 464-465.

Croll, J. 1889. *Stellar Evolution and Its Relation to Geological Time*, Edward Stanford, London.

Darwin, G. H. 1903. Radio-activity and the age of the Sun, *Nature* **68**: 496.

Darwin, G. H. 1905. [Presidential address], *Report, Brit. Assoc. Adv. Sci.* **75**: 3-32.

Daub, E. E. 1970. Maxwell's demon, *Stud. Hist. Phil. Sci.* **1**: 213-227.

Eddington, A. S. 1917. Further notes on the radiative equilibrium of the stars, *Month. Not. Royal Astron. Soc.* **77**: 596-612.

Eddington, A. S. 1920. The internal constitution of the stars, *Nature* **106**: 14-20.

Elkana, Y. 1974. *The Discovery of the Conservation of Energy*, Harvard University Press, Cambridge, MA.

Feulner, G. 2012. The faint young Sun problem, *Rev. Geophys.* **50**: RG2006.

Foster, J. B. and P. Burkett 2008. Classical Marxism and the second law of thermodynamics, *Org. Environ.* **21**: 3-37.

Giebeler, H. 1912. Spektroskopischer Beobachtungen der Nova Geminorum 2 am Bonner Refraktor, *Astron. Nachr.* **191**: 393-402.

Gold, B. J. 2010. *Thermopoetics: Energy in Victorian Literature and Science*, MIT Press, Cambridge, MA.

Helmholtz, H. 1995. *Science and Culture: Popular and Philosophical Essays*, ed. D. Cahan, University of Chicago Press, Chicago.

Herschel, J. 1833. *Treatise on Astronomy*, Green & Longman, London.

Hufbauer, K. 1981. Astronomers take up the stellar-energy problem, *Hist. Stud. Phys. Sci.* **11**: 277-303.

Hufbauer, K. 2006. Stellar structure and evolution, 1924-1939, *J. Hist. Astron.* **37**: 203-227.

Irwin, P. G. J. 2009. *Giant Planets of Our Solar System*, Springer, Berlin.

James, F. A. J. L. 1982. Thermodynamics and sources of solar heat, 1846-1862, *Brit. J. Hist. Sci.* **15**: 155-181.

James, Frank A. J. L. 1985. Between two scientific generations: John Herschel's rejection of the principle of the conservation of energy in his 1864 correspondence with William Thomson, *Notes Records Roy. Soc. London* **40**: 53-62.

Joly, J. 1903. Radium and the Sun's heat, *Nature* **68**: 572.

Jones, H. C. 1903. *A New Era in Chemistry*, Van Nostrand, New York.

Kant, I. 1981. *Universal Natural History and Theory of the Heavens*, ed. S. Jaki, Scottish Academic Press, Edinburgh.

Kant, I. 2015. *Natural Science*, ed. E. Watkins, Cambridge University Press, Cambridge.

Kayser, H. 1912. Ein Versuch zur Erklärung der neuen Sterne durch Radioaktive Prozesse, *Astron. Nachr.* **191**: 421-426.

Kidwell, P. A. 1981. Prelude to solar energy: Pouillet, Herschel, Forbes and the solar constant, *Ann. Sci.* **38**, 457-476.

Kragh, H. 2007. Cosmic radioactivity and the age of the universe, 1900-1930, *J. Hist. Astron.* **38**: 393-412.

Kragh, H. 2008. *Entropic Creation: Religious Contexts of Thermodynamics and Cosmology*, Ashgate, Aldershot.

Kragh, H. 2012. Is space flat? Nineteenth-century astronomy and non-Euclidean geometry, *J. Astron. Hist. Heritage* **15**: 149-158.

Kuhn, T. S. 1969. Energy conservation as an example of simultaneous discovery, pp. 321-356 in *Critical Problems in the History of Science*, ed. M. Clagett, University of Wisconsin Press, Madison.

Lane, J. H. 1870. On the theoretical temperature of the Sun, *Amer. J. Sci.* **50**: 57-84.

Lankester, E. R. 1906. [Presidential address], *Nature* **74**: 321-333.

Lindsay, R. B. 1973. *Julius Robert Mayer: Prophet of Energy*, Pergamon Press, Oxford.

Mayer, J. R. 1893. *Kleinere Schriften und Briefe*, ed. J. J. Weyrauch, Cotta'schen Buchhandlung, Stuttgart.

Mitchell, S. A. 1912. Radium and the chromosphere, *Astron. Nachr.* **192**: 266-270.

Mitchell, S. A. 1913. Is radium in the Sun? *Pop. Astron.* **21**: 321-331.

Moulton, F. R. 1914. *An Introduction to Celestial Mechanics*, Macmillan, New York.

Mousoutzanis, A. 2014. *Fin-de-Siécle Fictions, 1890s-1990s: Apocalypse, Technoscience, Empire*, Palgrave, London.

Newcomb, S. 1878. *Popular Astronomy*, Macmillan & Co, London.

Perry, J. 1895. On the age of the Earth, *Nature* **51**: 224-227.

Poincaré, H. 1911. *Leçons sur les Hypothèses Cosmogoniques*, A. Hermann, Paris.

Pouillet, C. S. 1838. Mémoire sur la chaleur solaire, sur les pouvoirs rayonnants et absorbants de l'air atmosphérique, et sur la température de l'espace, *Comptes Rendus* **7**: 24-65.

Powell, C. S. 1988. J. Homer Lane and the internal structure of the Sun, *J. Hist. Astron.* **19**: 183-199.

Rankine, W. M. 1881. *Miscellaneous Scientific Papers*, Griffin and Co., London.

Ritter, A. 1898. On the constitution of gaseous celestial bodies, *Astrophys. J.* **8**: 293-315.

Robitaille, P.-M. 2011. A thermodynamic history of the solar constitution, *Progress Phys.* **3**: 3-25, 41-59.

Rutherford, E. 1906. *Radioactive Transformations*, Cambridge University Press, Cambridge.

Rutherford, E. 1962. *The Collected Papers of Lord Rutherford of Nelson*, ed. J. Chadwick, vol. 1, Allen and Unwin, London.

Schaffer, S. 1995. Where experiments end: Tabletop trials in Victorian astronomy, pp. 257-299 in *Scientific Practice: Theories and Stories of Doing Physics*, ed. J. Z. Buchwald, University of Chicago Press, Chicago.

Shahiv, G. 2009. *The Life of Stars: The Controversial Inception and Emergence of the Theory of Stellar Structure*, Springer, Heidelberg.

Siemens, W. 1882. On the conservation of solar energy, *Proc. Roy. Soc.* **33**: 389-398.

Siemens, W. 1883. *On the Conservation of Solar Energy*, Macmillan, London.

Smith, C. and M. Norton Wise 1989. *Energy and Empire: A Biographical Study of Lord Kelvin*, Cambridge University Press, Cambridge.

Soddy, F. 1909. *The Interpretation of Radium*, John Murray, London.

Soddy, F. 1975. *Radioactivity and Atomic Theory*, ed. T. J. Trenn, Taylor & Francis, London.

Spencer, H. 1867. *First Principles*, Williams and Norgate, London.

Stark, J. 1903. [No title], *Nature* **68**: 230.

Stewart, B. 1870. What is energy? *Nature* **2**: 270-271.

Stinner, A. 2002. Calculating the age of the Earth and the Sun, *Phys. Educ.* **37**: 296-305.

Tait, P. G. 1876. *Lectures on Some Recent Advances in Physical Science*, Macmillan, London.

Tassoul, J.-L. and M. Tassoul 2004. *A Concise History of Solar and Stellar Physics*, Princeton University Press, Princeton.

Thomson, W. 1882. *Mathematical and Physical Papers*, vol. 1, Macmillan, London.

Thomson, W. 1884. *Mathematical and Physical Papers*, vol. 2, Macmillan, London.

Thomson, W. 1891. *Popular Lectures and Addresses*, vol. 1, Macmillan, London.

Tort, A. C. and F. Nogarol 2011. Another look at Helmholtz's model for the gravitational contraction of the Sun, *Europ. J. Phys.* **32**: 1679-1685.

Vogel, H. C., ed. 1892. *Newcomb-Engelmann's Populäre Astronomie*, W. Engelmann, Leipzig.

Waterston, J. J. 1860. On the inductions with respect to the heat engendered by the possible fall of a meteor into the Sun, *Month. Not. Roy. Astron. Soc.* **20**: 197-202.

Waterston, J. J. 1892. On the physics of media that are composed of free and perfectly elastic molecules in a state of motion, *Phil. Trans. Roy. Soc. A* **183**: 1-79.

Wesemael, F. 2009. Harkins, Perrin and the alternative paths to the solution of the stellar-energy problem, 1915-1923. *J. Hist. Astron.* **40**: 277-296.

Williams, W. M. 1870. *The Fuel of the Sun*, Simpkin, Marshall & Co., London.

Winnek, H. C. 1970. Science and morality in Thomas C. Chamberlin, *J. Hist. Ideas* **31**: 441-456.

Young, C. A. 1896. *The Sun*, Appleton and Company, New York.